\documentstyle[floats,twocolumn,amssymb,aps,epsfig,pre,color]{revtex}
\draft

\begin{document}
\wideabs{ \title{Stress transmission through three-dimensional granular crystals with 
stacking faults} \author{Melissa J. Spannuth, Nathan W. Mueggenburg,
Heinrich M. Jaeger, Sidney R. Nagel} \address{The James Franck
Institute and Department of Physics\\ The University of Chicago \\
5640 S. Ellis Ave. Chicago, IL 60637} \date \today \maketitle

\begin{abstract}
We explore the effect of stacking fault defects on the transmission of
forces in three-dimensional face-centered-cubic granular crystals.  An
external force is applied to a small area at the top surface of a
crystalline packing of granular beads containing one or two stacking
faults at various depths.  The response forces at the bottom surface
are measured and found to correspond to predictions based on vector
force balance within the geometry of the defects.  We identify the
elementary stacking fault as a boundary between two pure
face-centered-cubic crystals with different stacking orders.  Other
stacking faults produce response force patterns that can be viewed as
resulting from repetitions of this basic defect. As the number of
stacking faults increases, the intensity pattern evolves toward that
of an hexagonal-close-packed crystal.  This leads to the conclusion
that the force pattern of that crystal structure crystal can be viewed
as the extreme limit of a face-centered-cubic crystal with a stacking
fault at every layer.
\end{abstract}

}

\section{Introduction}

Static packings of granular particles support stress in a nontrivial
manner \cite{Jaeger96}.  The forces on any given particle from all
neighboring particles and from gravity must balance vectorally.
Stress on a random granular pack is supported by a disordered network
of these interparticle contacts in local force balance with each other
\cite{Liu95,Mueth98}.  It has been suggested that elastic theory is
capable of accounting for these inhomogenaities
\cite{Goldenberg01,Goldhirsch02}.  However, the indeterminacy when the
particles are infinitely hard, so that particle distortions are not
allowed, has led to a search for alternative descriptions of the
mechanical properties of granular materials in the hard-sphere limit
\cite{Bouchaud95,Wittmer96,Edwards98,Head01}.

The proposed alternatives have had varying degrees of success.
Coppersmith {\em et al.} proposed a diffusive q-model for the
transmission of force through granular packings
\cite{Liu95,Coppersmith96}, and a hyperbolic model was created by
Cates {\em et al.} \cite{Cates98}.  Experimentally it has been
difficult to rule out any of these descriptions conclusively, but
recently progress has been made in studying the conditions under which
these models apply.

Much understanding and additional surprises have been gained from
experiments studying the response to a localized external force,
similar to a Green's function.  DaSilva and Rajchenbach
\cite{DaSilva00} studied a two dimensional packing of rectangular
photoelastic bricks.  When applying an external force at the top of
the packing they measured the response forces between neighboring
bricks within the pack.  The results showed that at a given depth the
maximum interparticle force was located directly below the position of
the applied force.  Deeper in the packing (farther from the applied
force) they found the width of this peak of maximum force to broaden
as the square root of the depth.  Similar results were found by
Moukarzel {\em et al.} in recent displacement response experiments
\cite{Moukarzel03}.  Such findings support diffusive models.  However,
a cross-over to a linear widening of the response peak
cannot be ruled out for deeper packings.  Reydellet and Cl\'ement looked
at the response forces at the bottom surface of an amorphous
three-dimensional packing of spheres \cite{Reydellet01}.  Again they
found the maximum force situated directly beneath the applied force,
but now the width of the peak grew linearly with depth from the
surface.  This linear growth lends credence to the elastic theories.
Similar elastic-type behavior was found in two-dimensional packings by
Geng {\em et al.}.  In addition those researchers also found a strong
dependence on the spatial ordering of the particles \cite{Geng00,Geng03}.

The importance of spatial ordering was further shown in three
dimensional granular crystals in Ref. \cite{Mueggenburg02}.
Within face-centered-cubic (FCC) crystals, the force was found to be
supported along straight lines of contacts between beads, resulting in
three areas of large force at the bottom surface in response to a
locally applied external load at the top surface.  In contrast, an
hexagonal-close-packed (HCP) arrangement resulted in a ring of large
force at the bottom surface.  In both cases there was a local minimum
of force directly beneath the point of application.  Such patterns are
reminiscent of the hyperbolic models, which predict straight-line
propagation of forces.  Bouchaud {\em et al.} have extended these
hyperbolic models to include the splitting of force chains at defects
within a granular packing \cite{Bouchaud00}.  Such splittings are
predicted to lead to the elastic-like behavior that has been observed
in amorphous packings.

It is, therefore, of interest to explore the regime between
perfectly-ordered crystalline bead packs and completely amorphous
arrangements.  As a first step in this direction, we present an
experimental study of the effect on the force propagation of a small
number of stacking fault defects within otherwise perfectly ordered
FCC crystals.

\section{Experimental Methods}

Employing the triangular acrylic cell described by Blair {\em et al.}
\cite{Blair01} and Mueggenburg {\em et al.} \cite{Mueggenburg02} and a
'by hand' construction technique, we produced large FCC crystals of
approximately $20,000$ soda lime glass spheres of diameter $3.06 \pm
0.04$mm with an extremely low number of defects (estimated to be fewer
than ten beads significantly out of place in the entire pack) oriented
as horizontal planes of triangular order. Furthermore, we carefully
controlled the stacking of the planes in order to create true
three-dimensional crystals and to introduce stacking faults at
specific locations.

In this manner, we constructed five different granular FCC crystals,
each having one or two defects in the stacking order.  We then applied an
impulsive force to a small region, approximately two beads in diameter,
centered on the top of the crystal \cite{Mueggenburg02}.  Using a
carbon paper technique, we characterized the normal forces on the top
and bottom surfaces of the crystal in response to this force
\cite{Liu95,Mueth98,Mueggenburg02,Delyon90,Blair01}.  A piece of
carbon paper was sandwiched between the crystal and a piece of white
paper on both the top and bottom surfaces of the crystal.  The applied
force caused individual beads to press into the carbon paper and leave
marks on the white paper.  These marks were then digitized and image
analysis software was used to calculate the area and intensity of each
mark \cite{Mueggenburg02}.

We conducted twenty experimental runs on each crystal.  For each run
the resultant carbon marks were aligned with the expected triangular
lattice.  The areas and intensities were averaged over experimental
runs and, in order to improve statistics, were averaged over
symmetries of the crystal (one reflection and two rotations) when such
averaging did not qualitatively alter the intensitiy patterns.

\section{Results}

A pure FCC crystal consists of layers of triangular order stacked in
such a way that every third layer lies on top of the first.  We
annotate this stacking as $\underline{ABCABCABCABCABC...}$  A stacking
fault corresponds to a disruption of this order and may be considered
to be a boundary between two pure FCC crystals with different stacking
orders.

\begin{figure}[tb]
\centerline{\epsfxsize=7.0cm\epsfbox{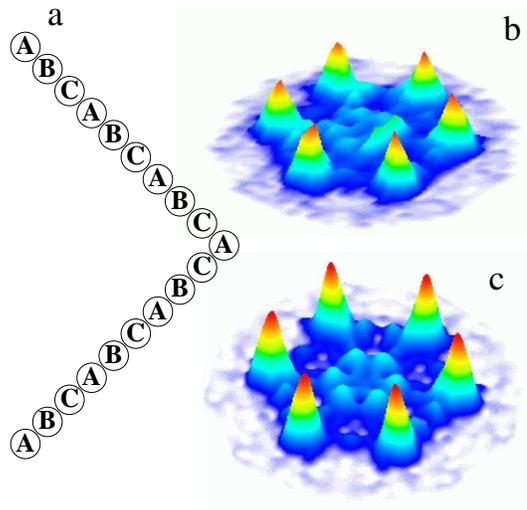}}
\vspace{.5cm}
\caption{19 layer FCC crystal with one stacking fault at layer 10.  a)
Schematic representation of layer stacking showing a boundary between
two FCC crystals.  b)  Force intensity plot at bottom surface in
response to a quick impulse applied to a small region at the center of
the top of the packing.  c)  Force intensity plot after averaging over
crystal symmetries, which accentuates anomalous interior features.}
\label{one_fault}
\end{figure}

We constructed a $19$ layer crystal with a stacking fault centered at
the tenth layer.  This stacking is represented as
$\underline{ABCABCABC}\underline{\overline{A}}\overline{CBACBACBA}$
The underline represents one pure FCC crystal, while the overline
corresponds to a second pure FCC crystal with a different orientation.
At the boundary between the two crystals there exists an
$\underline{\overline{A}}$ layer which correctly matches both stacking
orders and can thus be considered to be in both crystals.  Note that,
by necessity, the three layers centered around the stacking fault
$\underline{C}\underline{\overline{A}}\overline{C}$ follow an
hexgonal-close-packed stacking order.  When applying a quick impulse
to a small area at the center of the top of this crystal, the bottom
surface displayed six regions of large force at the vertices of a
regular hexagon as shown in figure \ref{one_fault}.  The regions of
large force are approximately the same size as the area over which the
impulse was applied on the top surface.  We observed another region of
larger than background force inside the hexagon, but closer to one
face.  At other symmetric places of the pattern, areas of larger than
background force are present, but with much lower intensity.  It is
not clear if this area of heightened force is a real feature of the
crystal with stacking fault or if this is merely an extreme
fluctuation.  Averaging over crystal symmetries would accentuate such
fluctuations, and thus we also show the intensity pattern without this
averaging.

\begin{figure}[tb]
\centerline{\epsfxsize= 6.0cm\epsfbox{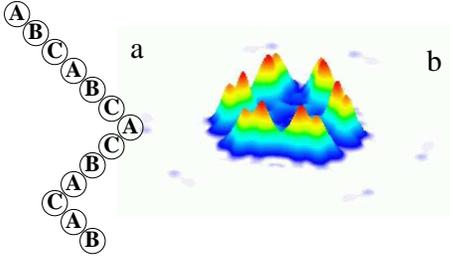}}
\vspace{.5cm}
\caption{a) Schematic representation of layer stacking of a 13 layer
crystal with faults centered at layers seven and eleven.  b)
Intensity plot of forces on the bottom surface showing twelve peaks of
high force.}
\label{two_faults_1}
\end{figure}

Additionally, we explored possible assemblies of crystals with two
stacking faults.  Figure \ref{two_faults_1} represents a $13$ layer
FCC crystal with stacking faults at the seventh and eleventh layers.
The stacking order (from top to bottom) was
$\underline{ABCABC}\underline{\overline{A}}\overline{CBA}\underline{\overline{C}}\underline{AB}$
as shown schematically in figure \ref{two_faults_1}(a).  We found that
this crystal produced a twelve-peaked intensity plot (Figure
\ref{two_faults_1}b).  The plot shows two pairs of peaks of nearly
equal height aligned along the faces of an equilateral triangle.  Each
pair of peaks is approximately four beads wide and the separation
between pairs is approximately three beads.

\begin{figure}[tb]
\centerline{\epsfxsize= 6.0cm\epsfbox{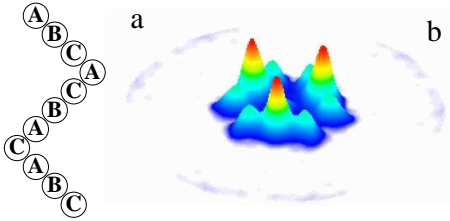}}
\vspace{.5cm}
\caption{a) Schematic representation of layer stacking of an 11 layer
crystal with faults centered at layers four and eight.  b)  Intensity
plot of forces on the bottom surface showing three large peaks each
flanked by a pair of smaller peaks.}
\label{two_faults_2}
\end{figure}

Another crystal with two stacking faults consisted of eleven layers
with faults centered at the fourth and eighth layers
($\underline{ABC}\underline{\overline{A}}\overline{CBA}\underline{\overline{C}}\underline{ABC}$).
For this crystal, we found three tall peaks flanked by peaks half the
height of the taller ones arranged along the faces of an equilateral
triangle (Figure \ref{two_faults_2}).  The width of each of the taller
peaks was slightly larger than that of the smaller flanking peaks
which correspond to the width of the area over which the force was
applied at the top surface.

\begin{figure}[tb]
\centerline{\epsfxsize= 6.0cm\epsfbox{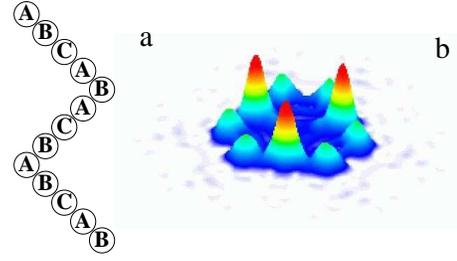}}
\vspace{.5cm}
\caption{a) Representation of layer stacking for a $13$ layer crystal
with stacking faults at the fifth and ninth layers.  b)  Intensity
plot of forces on the bottom surface showing three larger peaks each
flanked by a pair of smaller peaks.}
\label{two_faults_3}
\end{figure}

The next crystal we built contained thirteen layers with faults at the
fifth and ninth layers.  The stacking order was
$\underline{ABCA}\underline{\overline{B}}\overline{ACB}\underline{\overline{A}}\underline{BCAB}$.
We found an intensity pattern (Figure \ref{two_faults_3}) similar
to that of the previous crystal with two faults.  This crystal
produced three tall peaks flanked by peaks about one third the height
of the taller peaks arranged along the faces of an equilateral
triangle.  The taller peaks were approximately three beads wide and
the shorter peaks were approximately two beads wide.  In contrast to
figure \ref{two_faults_2}, which corresponds to a crystal with fewer
layers above and below the stacking faults, the flanking peaks are now
farther away from the taller peaks.

\begin{figure}[tb]
\centerline{\epsfxsize= 6.0cm\epsfbox{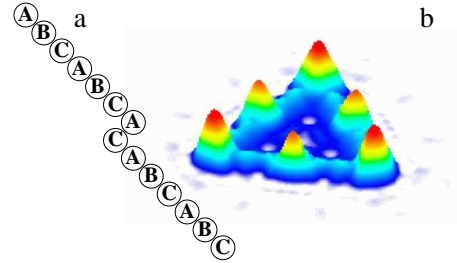}}
\vspace{.5cm}
\caption{a)  Layer stacking for a $14$ layer crystal with one missing
layer fault - also described as two grain boundary stacking faults at
layers seven and eight.  b)  Intensity plot of forces on the bottom
surface showing six peaks in a triangular arrangement.}
\label{two_faults_4}
\end{figure}

Another possible type of stacking fault involves the omission of one
plane from a large FCC crystal.  This fault maintains the orientation
of the crystal above and below the fault and thus might be considered
a more basic defect than the grain boundary described above.  However,
this fault could be described as a special case of two
grain-boundary faults.  Figure \ref{two_faults_4} shows the results
for a crystal with layer stacking
$\underline{ABCABC}\overline{\underline{A}\thinspace}\underline{\overline{C}}\underline{ABCABC}$.
Although one might suppose that this crystal exhibited the least
amount of stacking disorder, we found that the force intensity pattern
consisted of six approximately equal-height peaks, one at each of the
vertices of an equilateral triangle and one at each of the midpoints
of the sides (Figure \ref{two_faults_4}).  The peaks at the
vertices of the triangular pattern are four beads in width while those
on the sides are three beads wide.  We describe this pattern below as
a special case of the nine or twelve peaks shown in the crystal
packings with two faults.

\section{Discussion}

As in the experiments performed on pure FCC crystals by Mueggenburg
{\em et al.} \cite{Mueggenburg02}, we can reconstruct the force chain
paths through the crystal based on the force intensity plots from the
experiments.  More importantly, the simple force balance model used to
explain the force patterns in that study, also explains the present
results.

\begin{figure}[p]
\centerline{\epsfxsize= 6.0cm\epsfbox{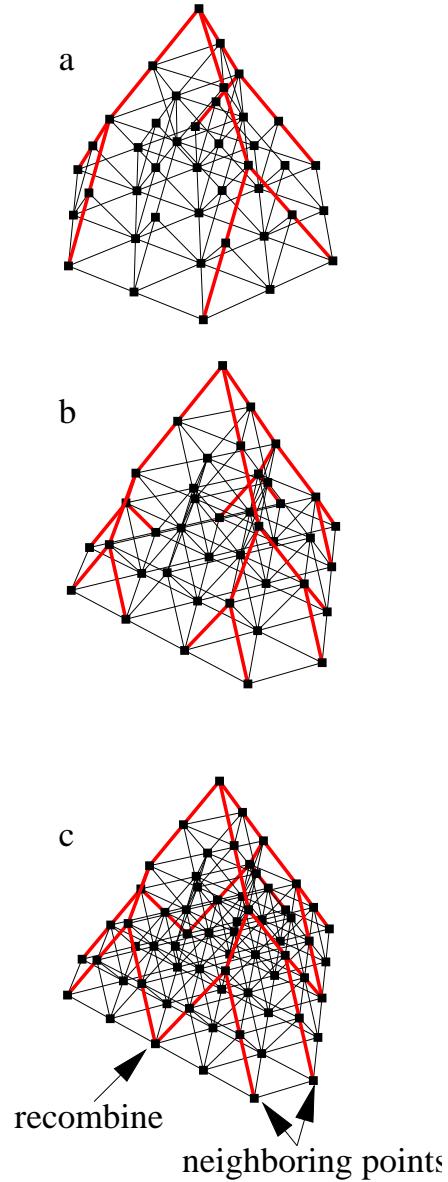}}
\vspace{.5cm}
\caption{Diagrams of force transmission through a crystal packing with
stacking faults.  a)  Forces must split at the stacking fault.  b)
Multiple stacking faults cause multiple force splittings.  c)  In some
cases multiple force chains may recombine or otherwise arrive at the
bottom surface at neighboring points and thus produce fewer areas of
large force.}
\label{diagrams}
\end{figure}

Forces will travel through pure FCC sections of the crystal, along the
spines of a tetrahedron.  Thus, to understand an elementary stacking
fault, all we need to determine is what happens at the interface
between two FCC crystals with different orientations.  Again, the
force balance model can be applied.  In order to maintain force
balance at the interface, the force chain must split.  Note that
locally this is the same geometry as an HCP crystal.  Below the
stacking fault each of the new forces will travel through the lower
portion of the crystal in straight line paths.  Figure \ref{diagrams}(a)
illustrates this force splitting at a grain boundary.

We found a regular hexagonal pattern of force in our first crystal
because the fault was placed in the middle of the packing.  In general,
we would expect a six-peaked pattern in which two peaks reside along
each side of a triangle.  As the defect is moved higher or lower in
the crystal, these pairs of peaks would move towards or away from each
other.

It is simple to extend this idea to crystals containing two stacking
faults.  The crystal can be thought of as three pure FCC crystals with
top and bottom sections having the same stacking order and the middle
section having a different stacking order.  At each interface the
force splitting process is repeated resulting in twelve peaks in the
intensity pattern (3 split into 6 and then split into 12).  Figure
\ref{diagrams}(b) illustrates this multiple splitting process.  However,
most of our results do not demonstrate the expected twelve-peaked
pattern.  This can be explained by force recombination.

For certain combinations of stacking faults, the forces will converge
upon the same lattice point on the bottom of the crystal.  In other
cases, two forces could travel along parallel lines through the
crystal and arrive at neighboring lattice points on the bottom
surface.  In the former scenario, the force intensity at the
convergence point should be twice the intensity of a lattice point
with only one line of force.  In the latter scenario, the forces
should produce a wider than normal peak (the exact width would depend
upon the width of the applied impulse and the separation of the force
chains within the packing).  These two effects account for our
observed force patterns from crystals with two faults.

In the second and third crystals with two stacking faults (Figures
\ref{two_faults_2} and \ref{two_faults_3}), the taller central peaks
are places where forces have recombined and the smaller flanking peaks
correspond to single force chains.  The widths of the peaks
corroborate this explanation in both cases.  The fourth example, where
one plane of the lattice has been removed, can be categorized as a
crystal with two stacking faults in which the twelve peaks have been
reduced to six due to this recombination.  The corner peaks in the
intensity plot (Figure \ref{two_faults_4}) are caused by parallel
forces terminating at adjacent lattice points and the side peaks are
caused by two forces terminating at the same lattice point as shown in
figure \ref{diagrams}(c).  The widths of the peaks support this
explanation.  This is the extreme case of the trend from figure
\ref{two_faults_2} to figure \ref{two_faults_3} where the flanking
peaks moved towards the corners of the triangular pattern as the
spacing between defects relative to the overall crystal height
decreased.  In the case, where the two defects occur at successive
layers, the flanking peaks are positioned very near the corners and
thus overlap with each other.

\section{conclusions}

In this study, we have established how stacking faults affect the
transmission of stress through an FCC crystalline granular assembly.
We have observed that the force chains split into two equal components
at the boundary between different FCC crystal orientations.  However,
we note that the forces are transmitted as in a pure FCC crystal
before and after splitting.  Our findings reinforce those in reference
\cite{Mueggenburg02} and the force balance model proposed therein.

We have identified the elementary stacking fault defect as the
boundary between two pure FCC crystals of different stacking orders.
All further stacking faults can be described as repetitions of this
basic defect.  Alternately, we can express the stacking faults as
sections of HCP crystals within the overall FCC crystal.  Although we
find the grain-boundary description the most lucid, this other
interpretation is geometrically equivalent.

As the number of stacking faults increases, the intensity pattern
looks increasingly like that of an HCP crystal, and indeed the local
stacking order around a fault is HCP.  This leads to the conclusion
that the force pattern of an HCP crystal can be viewed as the extreme
limit of an FCC crystal with a stacking fault at every layer.

\section{Acknowledgments}
This work was supported by NSF-CTS 0090490 and by the NSF MRSEC
Program under DMR-0213745. MJS acknowledges support by the University
of Chicago MRSEC Summer 2002 REU program.

\end{document}